\documentclass[aps,pra,reprint,superscriptaddress,nofootinbib,floatfix]{revtex4-2}

\usepackage{graphicx,xcolor}
\usepackage[colorlinks=true,
hyperfootnotes=true,
breaklinks=true,
citecolor=red,
urlcolor=blue,
linkcolor=blue
]{hyperref}

\begin{document}
\title{Hybrid discrimination strategy in quantum communication based on photon-number-resolving detectors and mesoscopic twin-beam states}
\author{Luca Razzoli}
\altaffiliation[Currently at ]{Department of Physics ``Alessandro Volta'', University of Pavia, Via Bassi 6, I-27100 Pavia (Italy); INFN, Sezione di Pavia, Via Bassi 6, I-27100 Pavia (Italy).}
\affiliation{Center for Nonlinear and Complex Systems, Department of Science and High Technology, University of Insubria, Via Valleggio 11, I-22100 Como (Italy)}
\affiliation{INFN, Sezione di Milano, Via Celoria 16, I-20133 Milano (Italy)}

\author{Alex Pozzoli}
\affiliation{Como Lake Institute of Photonics, Department of Science and High Technology, University of Insubria, Via Valleggio 11, I-22100 Como (Italy)}

\author{Alessia Allevi}
\email{alessia.allevi@uninsubria.it}
\affiliation{Como Lake Institute of Photonics, Department of Science and High Technology, University of Insubria, Via Valleggio 11, I-22100 Como (Italy)}

\date{\today}

\begin{abstract}
State discrimination is a key challenge in the implementation of quantum communication protocols. Most optical communication protocols rely on either coherent states of light or fragile single-photon states, making it often difficult to achieve robustness and security simultaneously. In this work, we propose a hybrid strategy that operates in the mesoscopic intensity regime, leveraging robust quantum states of light. Our approach combines classical and quantum features: reliable state discrimination based on a classical property of light, and security stemming from nonclassical correlations. Specifically, the receiver uses photon-number-resolving detectors to access the mean photon number of the binary thermal signals encoding the information. The communication channel exploits twin-beam states, inherently sensitive to eavesdropping attacks, to provide a layer of security. This strategy is scalable, allowing for straightforward extension to more complex signal alphabets, and offers a promising route for robust and secure quantum communication in the mesoscopic intensity domain.
\end{abstract}
\maketitle

\section{Introduction}\label{sec_intro}
In communication protocols, the discrimination of the alphabet in which a message is encoded represents a fundamental step \cite{cariolaro}. Depending on the complexity of the message and the security of the transmission channel, different strategies can be employed. In particular, the chosen alphabet could involve either discrete \cite{flamini} or continuous variables \cite{grosshans}. Usually, the main requirement is to develop a receiver that is able to discriminate the sent signals by minimizing the error probability in the discrimination process. To this aim, both direct and homodyne detection systems are commonly used \cite{diamanti,olivares}. More recently, hybrid detection systems have also been developed and successfully employed \cite{dimario1,dimario2,oe24,pla25}. In such schemes, photon-number-resolving (PNR) detectors are embedded in interferometric setups, enabling the use of both discrete and continuous variables.

In most cases, optical communication protocols involve coherent states of light since they are more robust than nonclassical states \cite{bergou}. However, the use of quantum resources could improve the security of the transmission channels by enabling the detection of eavesdropping attempts \cite{benenti}.
Actually, the main limitation in fully leveraging quantum states lies in their fragility. Indeed, at the single-photon level losses can significantly reduce the rate at which information is sent. Instead, in the mesoscopic intensity domain the optical states contain larger numbers of photons, thus being more robust against external degradation, while still exhibiting the nonclassical properties essential to the successful implementation of the protocol \cite{josab19,oe21}.
In this respect, here we introduce a communication protocol, whose transmission channel relies on the use of mesoscopic twin-beam (TWB) states and the alphabet adopted to encode information consists of two thermal signals with different mean values, superimposed on a twin portion of the above-mentioned TWB \cite{scirep22}.
The information thus encoded is transmitted by Alice, the sender, to Bob, whose detector is equipped with PNR capability. This choice allows for direct discrimination based on the mean value of light, assuming no eavesdropping occurs. However, in the case of an intercept-resend attack \cite{usenko}, where Eve intercepts a portion of light transmitted through the communication channel and re-sends a different signal with the same mean value, the discrimination does not guarantee the communication security. Nevertheless, exploiting nonclassical features can reveal the presence of an eavesdropping attack. In that case, the sequence under attack can be discarded and the communication interrupted. Specifically, we leverage the existence of nonclassical correlations in the number of photons of TWB states.
This enables Bob to implement a hybrid discrimination strategy, in which he exploits a classical property (the mean value of light) to discriminate and a nonclassical one (the noise reduction factor between the two arms of the TWB) to ensure security.

The reliability of the method is evaluated both in terms of error probability and by adopting figures of merit commonly used in binary classification, which provide insights into the type of error occurring within the communication channel. The investigation is performed as a function of the size of the data samples used to encode the message bits, a critical parameter that significantly impacts the performance of the system. Our results open new perspectives in the use of mesoscopic quantum resources to improve the security of communication channels.

\section{Theoretical framework}\label{sec_methods}

\subsection{Thermal states and mesoscopic TWB states}
The communication protocol we propose in this work is based on a binary alphabet, in which the bits 0 and 1 are encoded in two thermal states with different mean values, called noise signals hereafter. The choice leverages the PNR capability of the detection system, consisting of two Silicon photomultipliers (SiPMs) \cite{akindinov,agliati,ramilli}: Alice uses one, while Bob uses the other.
To be sufficiently general and to take into account the possible nonidealities in the experimental generation of the noise signals to be transmitted, we assume that such states are characterized by a multi-mode thermal distribution. In terms of detected photons, to which we have direct access, it can be expressed as \cite{mandel} 
\begin{equation}\label{multith}
P^{\mu}(m) = \frac{(m+ \mu -1)!}{m!(\mu-1)!(\langle m \rangle/\mu + 1)^{\mu}(\mu/\langle m \rangle +1)^m},
\end{equation}
where $\mu$ is the effective number of modes, while $\langle m \rangle$ is the mean number of detected photons.  
The photon-number statistics is super-Poissonian and can be also discriminated by means of the Fano factor. The latter is defined in terms of detected photons as $F(m) = \sigma^2(m)/\langle m \rangle= \eta F(n) +(1-\eta)$, where $\sigma^2(m)$ is the variance of the distribution of the number of detected photons and $\eta$ is the quantum efficiency of the detector \cite{arimondo}. It is well known that $F=1$ in the case of Poissonian light, $F<1$ for sub-Poissonian light (nonclassical), and $F>1$ for super-Poissonian light.

The two noise signals are sent in an alternative way through a communication channel together with a portion of a multi-mode TWB state \cite{pla22}, which is a quantum-correlated optical state. If the $\mu$ modes of each of the two parties of the TWB are equally populated, then the multi-mode TWB state is ideally described by a tensor product of $\mu$ identical TWB states \cite{epl10,silberhorn2016},
\begin{equation} \label{multiTWB}
\vert \Psi^{\mu}_{\rm TWB} \rangle = 
\bigotimes_{k=1}^\mu \sqrt{1- \lambda^2}
\sum_{\nu=0}^{\infty} \lambda^{\nu} \vert \nu_k \rangle \otimes \vert \nu_k\rangle,
\end{equation}
where $k$ labels the modes, $\nu$ is the photon number in each mode, $\lambda$ is defined through $\lambda^2 = \langle n \rangle / (\mu + \langle n \rangle)$, and $\langle n \rangle$ is the mean total number of photons in either of the two parties of the TWB, with $n=\sum_{k=1}^\mu \nu_k$. The state in Eq.~(\ref{multiTWB}) is characterized by perfect photon-number correlations between the two parties of the TWB in each mode $k$. It is possible to prove that the probability of observing an overall number of photons $n$ in each arm of the TWB is given by the multi-mode thermal photon-number distribution in Eq.~(\ref{multith}) written in terms of the number $n$ of incident photons \cite{paleari,mauerer,pla22}.
The measurements performed by Bob return a mean value that is given by the sum of the mean value of the thermal signal and that of the TWB portion. The other TWB portion is retained and measured by Alice herself. 
The presence of the TWB in the communication channel allows for revealing possible eavesdropping attacks. For instance, this can be achieved by applying nonclassicality criteria involving the number of photons, such as that based on the noise reduction factor, $R$. In terms of the number of detected photons, it reads as \cite{heidmann} 
\begin{equation} \label{Rdetphot}
R = \frac{\sigma^2(m_1-m_2)}{\langle m_1 \rangle + \langle m_2 \rangle},
\end{equation}
where $\sigma^2(m_1-m_2)$ is the variance of the distribution of the photon-number difference between the two parties of the TWB and $(\langle m_1 \rangle + \langle m_2 \rangle)$ is the shot-noise level. Values of $R$ lower than 1 prove that the states are sub-shot-noise correlated. It has been demonstrated that this is also a sufficient criterion for entanglement \cite{agliati}.
For the purpose of the protocol, it is useful to express the noise reduction factor in the presence of the thermal noise signal to be transmitted. To model a more realistic scenario, we also assume that the communication channel is affected by losses, described by the imbalance parameter $t$ between the two arms of the TWB \cite{oe21,pla22}:
\begin{eqnarray} \label{noisylossyR}
R &=& 1- \frac{2\eta t \langle m \rangle}{(1+t) \langle m \rangle + \langle m_{\rm N} \rangle} 
+ \frac{(1-t)^2 \langle m \rangle^2}{\mu \left[(1+t) \langle m \rangle + \langle m_{\rm N} \rangle \right]}\nonumber \\
&+& \frac{\langle m_{\rm N} \rangle^2}{\mu_{\rm N}[(1+t)\langle m \rangle + \langle m_{\rm N} \rangle]},
\end{eqnarray}
where $\langle m \rangle$ is the mean number of detected photons in a TWB arm, $\eta$ is the global quantum efficiency of the detection system (including the detection efficiency and the coupling of light into fibers), $t \in [0,1]$ is the transmission efficiency quantifying the balancing between the two arms, while $\langle m_{\rm N} \rangle$ and $\mu_{\rm N}$ are the mean value and the effective number of modes of the thermal noise signal, respectively.
This means that, in general, the value of $R$ increases in the presence of noise, since the latter is not correlated with the TWB, and also in the presence of losses, as $t<1$.
In this scenario, nonclassical correlations, $i.e.$ $R<1$, can be detected provided that
\begin{equation}
\langle m_{\rm N} \rangle < \sqrt{\mu_{\rm N} \left[ 2 \eta t \mu - (1-t)^2 \langle m \rangle \right]\langle m \rangle/\mu}.
\label{eq:Rless1_condition_mN}
\end{equation}
Physically, this inequality can be interpreted as follows: in the case of perfect balance between the TWB arms (\textit{i.e.}, for $t = 1$), the nonclassicality condition requires that the mean value of the signal noise, $\langle m_{\rm N}\rangle$, be less than a term proportional to $\sqrt{2\langle m \rangle}$, where $2 \langle m \rangle$ is the shot-noise level, corresponding to the total mean number of detected photons in the two TWB arms; for $t < 1$, the presence of an imbalance imposes a more stringent constraint on the thermal noise signal, requiring $\langle m_{\rm N}\rangle$ to be lower than that in the balanced case.

\subsection{Discrimination of the noise signal as a binary classification problem}
\label{sec_bin_class_pbm}
The binary alphabet adopted to encode the information comprises two noise signals, characterized by different mean values and encoding, respectively, the logical bits 0 and 1.
Letting $x$ be a scalar quantity that characterizes the noise signal and according to which we discriminate the latter, we label noise signals with values $x<x_{\rm th}$ ($x \geq x_{\rm th}$) as bit 0 (bit 1), where $x_{\rm th}$ is an arbitrary discrimination threshold value.
Discriminating between the two noise signals can be framed as a binary classification problem, whose performance can be assessed using various figures of merit, the most intuitive being the error probability, $P^{(\rm err)}$.
 Assuming that the two noise signals are characterized by equal prior probability, \textit{i.e.}, $p(0) = p(1) = 1/2$, $P^{(\rm err)}$ is expressed as~\cite{helstrom}
\begin{equation} \label{errprob}
P^{(\rm err)} = \frac{1}{2} [p(0|1) + p(1|0)],
\end{equation}
where $p(i|j)$ is the conditional probability of detecting $i$ having sent $j$.

To better characterize the performance of the binary classifier, we also consider the Receiver Operating Characteristic (ROC) curve \cite{roc_fawcett,roc_canbek}, a parametric plot that shows \textit{true positive rate} ${\rm TPR} = {\rm TP} / ({\rm TP+FN})$ against the \textit{false positive rate} ${\rm FPR} = {\rm FP}/ ({\rm FP+TN})$ at varying discrimination threshold values. By definition, ${\rm TPR},{\rm FPR} \in [0,1]$. 
In our setting, we treat bit 1 as the positive class and bit 0 as the negative. Thus, TP (true positives) and TN (true negatives) refer respectively to the number of correctly classified 1s and 0s, while FP (false positives) and FN (false negatives) to the number of 0s misclassified as 1s and of 1s misclassified as 0s, respectively.
For completeness, the false negative rate is ${\rm FNR = 1- TPR}$ and the true negative rate is ${\rm TNR = 1- FPR}$. Note that the number of T/F P/N, as well as the error probability in Eq. (\ref{errprob}), which can be expressed as $P^{\rm (err)} = ({\rm FNR}+{\rm FPR})/2$, are determined at fixed threshold value.
Based on these definitions, the perfect classifier corresponds to the point $({\rm FPR},{\rm TPR})=(0,1)$ (no false negatives and no false positives), while a random classifier is any point along the line ${\rm TPR}={\rm FPR}$. Accordingly, points above this diagonal represent better-than-random classifiers. Regarding the discrimination threshold, low threshold values correspond to points near $(1,1)$  (most instances are labeled as positives), while high threshold values correspond to points near $(0,0)$ (very few instances are labeled as positives).

The area under the ROC curve (AUC) score \cite{auroc} is a single, threshold-independent value which is often used to capture the overall performance of a binary classifier: ${\rm AUC} = 1$ for the perfect classifier and ${\rm AUC} = 0.5$ for a random one.

In the following, we address the binary discrimination based either on the mean value or the noise reduction factor and we investigate
the figures of merit introduced above for representative sizes of the sample used to encode a single logical bit.

\begin{figure*}[!t]
\centering
\includegraphics[width=1\textwidth]{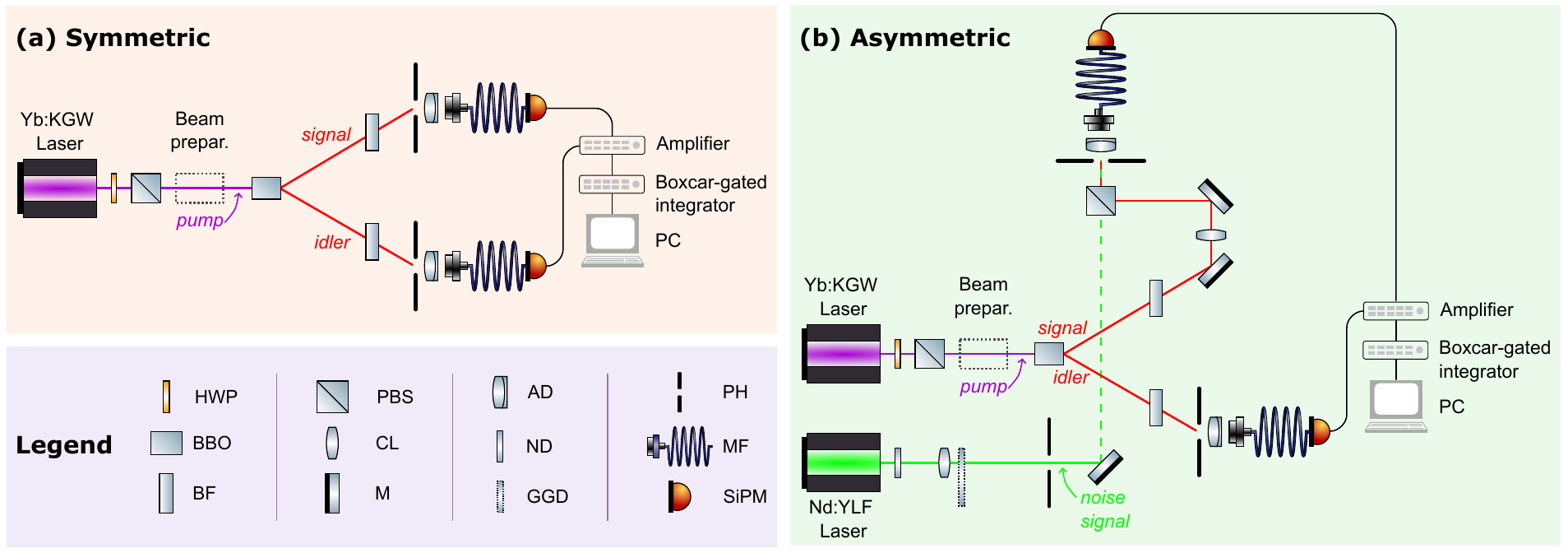}
\caption{Sketch of the experimental setup (a) for the characterization of TWB states and (b) for the communication protocol. HWP: half-wave plate; BBO: $\beta$-barium-borate crystal; BF: band-pass filter; PH: pin-hole; AD: achromatic doublet; MF: multi-mode optic fiber; SiPM: silicon photomultiplier; M: mirror; PBS: polarizing beam splitter; CL: convex lens; ND: neutral density filter; GGD: ground-glass disk. The stage of beam preparation (dashed box) comprises two prisms and a demagnifying telescope, as detailed in \cite{APL25}. The pump field is indicated in purple, the signal (idler) beam is shown as the upper (lower) red line, and in (b) the noise signal, generated by a different laser source, is indicated in green. The dashed line denotes that the noise signal is superimposed on the signal arm of the TWB in post-processing by suitably combining the acquired data.}
\label{setup}
\end{figure*}
\section{Experimental setup}\label{sec:exp_setup_twb}
\subsection{Communication channel}
The investigation of the communication protocol consists of two consecutive steps. First, we characterize the communication channel based on the multi-mode TWB state. In particular, we study both classical and nonclassical features of the TWB, such as the photon-number statistics and the strength of nonclassical correlations. Second, we investigate the performance of state discrimination superimposing the two noise signals on the portion of TWB sent to Bob.

The experimental setup we realized to accomplish these two steps is shown in Fig.~\ref{setup}: the sub-picosecond third-harmonic pulses at 344 nm of a tunable Yb:KGW laser are optically tilted by a system composed of two prisms and a demagnifying telescope before impinging on a $\beta$-barium-borate crystal (BBO), where they produce parametric down conversion. As described in \cite{APL25}, the tilting of the pulses is necessary to improve their group-velocity matching \cite{smith,schimpf} so as to obtain the phase-matching condition in the nonlinear crystal around frequency degeneracy (\textit{i.e.}, at 694 nm). The energy of the pump beam, and thus of the generated TWB, is changed by means of a half-wave plate (HWP) followed by a polarizing cube beam splitter (PBS) placed at the output of the laser.

Outside the BBO, spatial and spectral filtering (identical in the two arms) allows to select two entangled regions. As to the spectral filtering, we use two band-pass filters, having a central wavelength of 692 nm on one arm and 697 nm on the other arm, and both having a width of $\sim 20$ nm. For the spatial filtering, we use an adjustable pin-hole having a diameter of 1 cm. The filtered light in each arm is then focused by an achromatic doublet into a multi-mode optical fiber having a 1-mm-core diameter, and then delivered to the detector. The detection is performed by means of two SiPMs (model S13360-3050CS, Hamamatsu Photonics) \cite{sipm}. These are solid-state PNR detectors endowed with hundreds of cells (pixels), operated in the Geiger-M\"{u}ller regime. Under the assumption that each cell is fired by at most one photon impinging on it, each SiPM yields a single output that is proportional to the number of fired cells.
The SiPM model used in this work is characterized by a quantum efficiency $\sim 16 \%$ at 693 nm, which can reach higher values, $\sim 40 \%$, in the blue–green spectral region \cite{APL25}.
The detectors are plugged into a power supply and amplification unit (PSAU, model SP5600, CAEN) that supplies a bias voltage of 56 V to the SiPMs and amplifies the output by 17 dB. Each output is synchronously integrated over an interval of 10 ns by a boxcar-gated integrator and acquired.

\subsection{Binary noise signal}
The experimental setup for the study of the discrimination protocol is sketched in Fig. \ref{setup}(b). It is equivalent to the previous one [Fig. \ref{setup}(a)] except for the additional optical elements required to generate the noise signal and to superimpose it on the signal arm of the TWB. 
The thermal noise is generated by sending picosecond pulses at 523 nm to a ground-glass disk. A single speckle is selected in far field by an iris 300-$\mu$m large. The energy of the laser beam is changed by means of a variable neutral density filter.
In this way, it is possible to produce the two noise signals with different mean values, necessary to encode the logical bits 0 and 1, that can be superimposed in an alternative way on the portion of TWB sent to Bob. In the present study, we superimpose the noise signals in post-processing [see the dashed line in Fig.~\ref{setup}(b)], following a similar procedure already adopted in \cite{perina,pla23} to merge the data corresponding to TWB states and those corresponding to noise signals.
In a practical realization of the protocol, the superposition of one of the two noise signals on the signal arm of TWB could represent a critical point. In fact, even though an imperfect overlap, due for instance to mode matching, could be included in the expression in Eq.~(\ref{noisylossyR}), its presence could make the two signals more distinguishable, thus degrading the security of the protocol \cite{usenko0}. However, it has been demonstrated that this issue could be mitigated through the use of a homodyne scheme with a strong local oscillator \cite{usenko1} or by employing non-Gaussian modulations \cite{usenko2}.
Moreover, superimposing two beams by means of optical elements, such as beam splitters and dichroic mirrors, could introduce further losses, because both the signal of TWB and the thermal states could be attenuated. However, this effect is not particularly detrimental since it can be included in the parameter $t$ for the TWB and would act in the same way for the two thermal states at different mean values. To take into account the role played by $t < 1$ in the practical implementation of the communication channel based on TWB, we decide to insert in the setup some optical elements that increase its length and introduce some losses. Despite the latter, nonclassical correlations can still be detected by Bob, thus enabling the devised communication protocol.

\section{Results}\label{sec_results}
\subsection{Measurements and data processing}
The quantities we are interested in---the mean number of detected photons $\langle m \rangle$ and the noise reduction factor $R$---ultimately depend on the intensity of the light that impinges on the two detectors. Therefore, it is necessary to preliminarily characterize the TWB states for different values of their intensity in the mesoscopic intensity domain, and the two noise signals we superimpose on the signal arm to encode the bits. Concerning the first aspect, we use a combination of HWP and PBS [Fig. \ref{setup}(a)] as a knob to tune the intensity of the pump beam and so, ultimately, to change the mean value of TWB. In more detail, we consider 4 angle values of the HWP placed on the pump beam path to investigate the statistical properties of the TWB alone as a function of the pump beam intensity. The TWB is then investigated in the presence of losses by using the asymmetric setup in Fig.~\ref{setup}(b). The two noise signals are also preliminarily characterized in order to verify that we have at our disposal single-mode thermal states. Their mean values are properly selected in order to satisfy Eq.~(\ref{eq:Rless1_condition_mN}), thus ensuring $R<1$.

\begin{figure}[!t]
\centering
\includegraphics[width=\columnwidth]{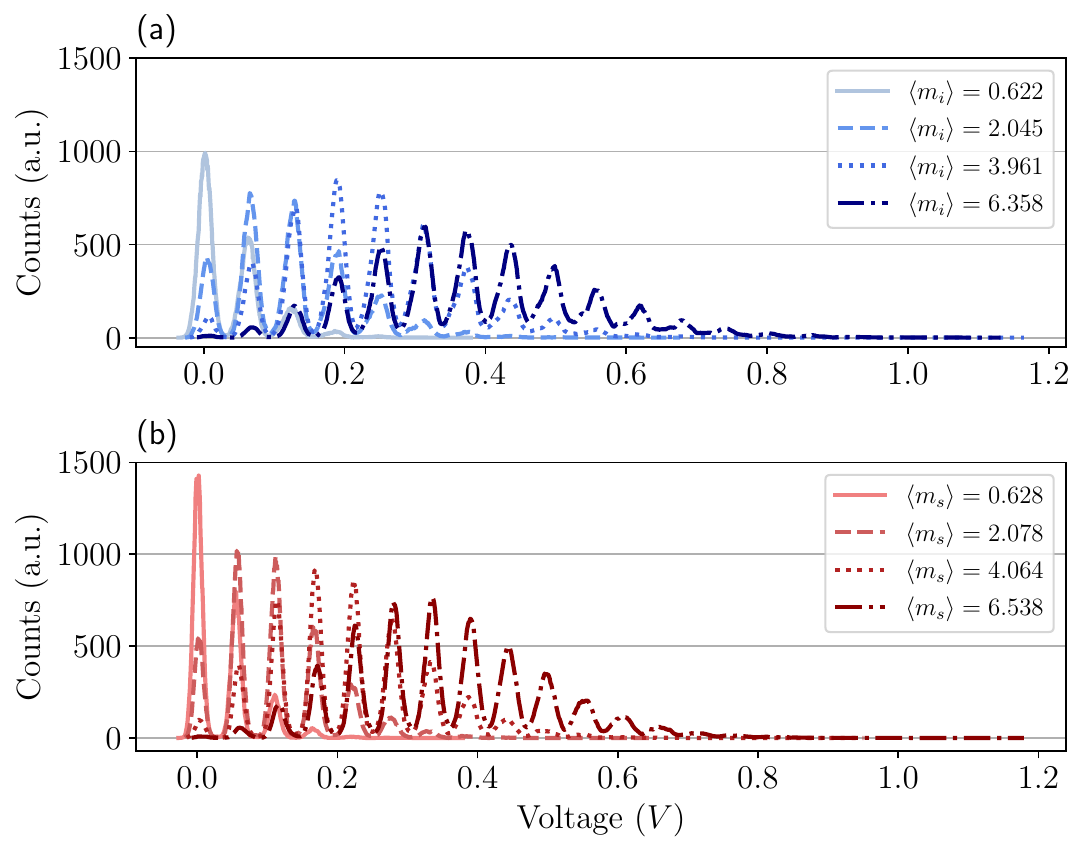}
\caption{Pulse-height spectra of idler (a) and signal (b) at different mean numbers of detected photons $\langle m \rangle$. Results are obtained from $10^5$ data for each mean value.}
\label{fig:twb_counts}
\end{figure}

Following the analysis procedure previously described in \cite{jmo}, for all the considered configurations the two outputs of the detection chain are suitably processed to reconstruct the statistical properties of light in terms of detected photons.
Specifically, we process the data acquired by each SiPM in every conditions by computing the pulse-height spectrum \cite{cassina} (see, \textit{e.g.}, Fig. \ref{fig:twb_counts}). From the latter we determine the gain of each detector as the average distance between the peaks of the histograms. Finally, the amplitude is converted from Volts (units of the measured outputs) to number of detected photons by dividing each column of data by the corresponding gain and rounding the results to the nearest integer value.

\subsection{Characterization of the communication channel}
\label{sec:results_twb}
\begin{figure}[!t]
\centering
\includegraphics[width=\columnwidth]{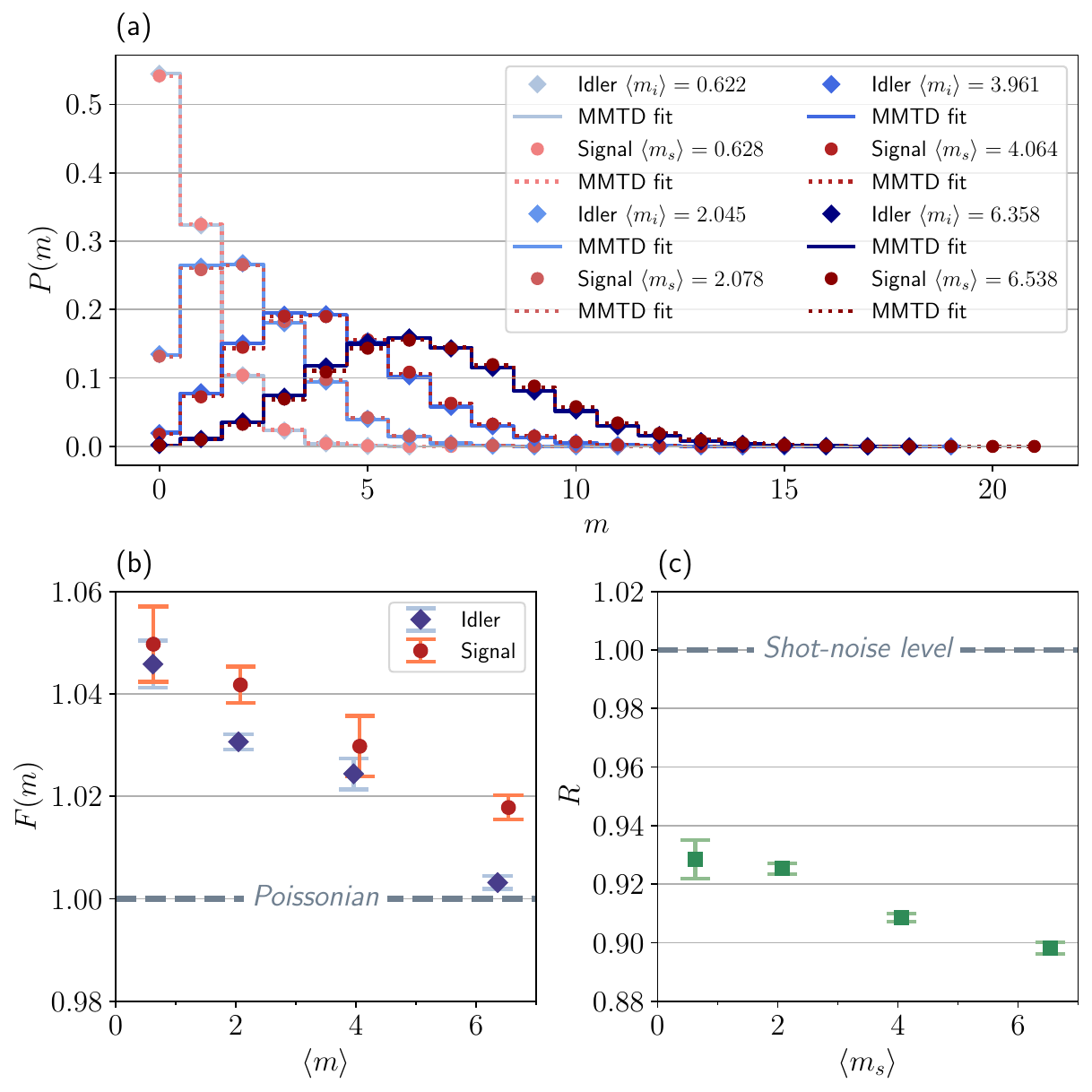}
\caption{Characterization of the symmetric TWB. (a) Detected-photon distribution $P(m_k)$ in the idler $k=i$ (blue-scale-colored diamonds) and signal $k=s$ (red-scale-colored circles). The fitting function (line) is the multi-mode thermal distribution (MMTD) in Eq. (\ref{multith}). Light (dark) colors correspond to low (high) $\langle m_k \rangle$. (b) Fano factor $F(m_k)>1$ of each arm as a function of $\langle m_k \rangle$. (c) Noise reduction factor $R<1$ as a function of $\langle m_s \rangle$. Results in (b) and (c) are obtained by dividing the $10^5$ data of each dataset into 4 batches of 25000 shots: the marker with its error bar represents the mean and the associated standard error computed from the batches of the considered dataset.}
\label{fig:twb_mmtd}
\end{figure}

As explained in Sec. \ref{sec_methods}, many properties of the TWB that qualify the communication channel can be determined starting from the number of detected photons $m_k$ in each arm, signal ($k=s$) and idler ($k=i$). Among them, here we consider the photon-number distribution, the Fano factor, and the noise reduction factor.
For all the four increasing values of the pump intensity, our results are consistent with the expected behavior of the TWB. The two arms are well balanced, $\langle m_s \rangle \approx \langle m_i \rangle$, and each detected-photon distribution is well fitted by the multi-mode thermal distribution in Eq.~(\ref{multith}), as shown in Fig. \ref{fig:twb_mmtd}(a).
The super-Poissonian nature of the measured light is also evidenced by the Fano factor, which is larger than 1 for all the datasets shown in Fig. \ref{fig:twb_mmtd}(b). Finally, the noise reduction factor is $R<1$ [see Fig. \ref{fig:twb_mmtd}(c)] in all cases, denoting the presence of nonclassical photon-number correlations in the mesoscopic intensity domain. 
By fitting the latter results using Eq. (\ref{noisylossyR}) with $t=1$ (lossless configuration, the two arms are balanced) and $\langle m_{\rm N} \rangle = 0$ (noiseless configuration), we can estimate the detection efficiency $\eta = (8.5 \pm 0.7)\%$, which is lower than the expected one ($\eta \sim 16\%$ at 693 nm, according to the datasheet\cite{sipm}). The discrepancy between our estimate and the expected value is likely due to possible losses in the collection of light including a non perfect coupling into fibers.
Remarkably, despite its low value, this efficiency is still sufficient to detect nonclassical correlations.

\begin{figure}[!t]
\centering
\includegraphics[width=\columnwidth]{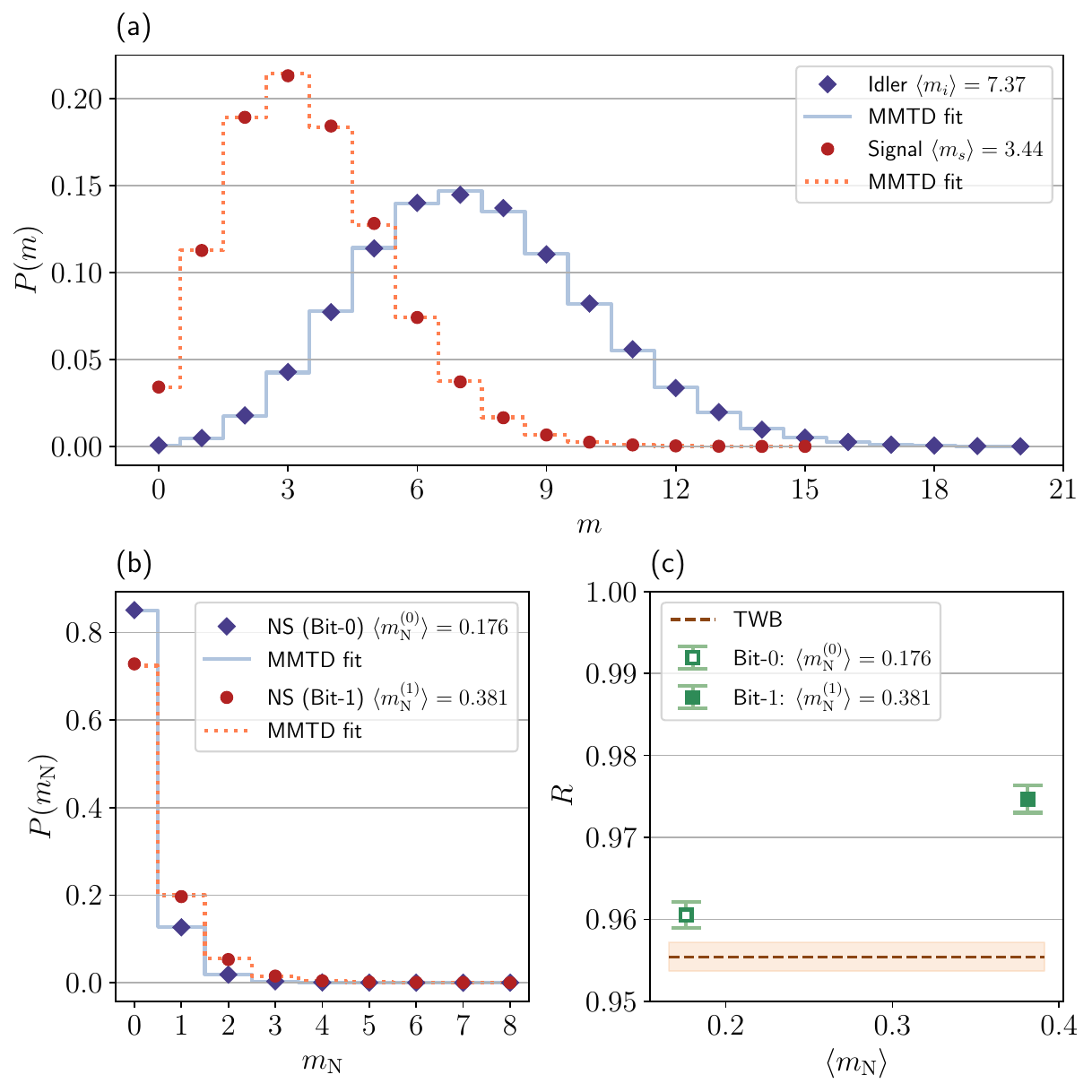}
\caption{Characterization of the asymmetric TWB and the noise signals. (a) Detected-photon distribution $P(m_k)$ in the idler $k=i$ and signal $k=s$ and (b) in the two noise signals (NS). The fitting function (line) is the multi-mode thermal distribution (MMTD) in Eq. (\ref{multith}). (c) Noise reduction factor of the TWB with superimposed noise signal as a function of $\langle m_{\rm N} \rangle$. The dashed line indicates the value of $R$ with associated error in the noiseless lossy configuration. Results in (c) are obtained using the same procedure described in Fig. \ref{fig:twb_mmtd}(c).}
\label{Rwithnoise}
\end{figure}

Due to the additional optical elements required to mimic the losses in the communication channel, in the configuration shown in Fig.~\ref{setup}(b) the TWB is no longer symmetric: The optical length of the signal has increased, while that of the idler is almost unchanged. These differences determine the presence of losses, especially in the signal arm. 
Losses can be easily estimated in terms of the transmission efficiency $t = \langle m_s \rangle / \langle m_i \rangle \approx 47\%$, where the mean number of detected photons in the signal and idler arms is $\langle m_s \rangle = 3.44 \pm 0.01$ and $\langle m_i \rangle = 7.37 \pm 0.01$, respectively.
Again, in each arm the detected-photon distribution is well described by the multi-mode thermal distribution in Eq.~(\ref{multith}) (see Fig.~\ref{Rwithnoise}(a), where the asymmetry introduced by the transmission channel is evident), with Fano factor $F(m_i)=(1.012 \pm 0.002)>1$ and $F(m_s)=(1.036 \pm 0.005)>1$, respectively. Despite the losses, we can still detect sub-shot-noise correlations.
Indeed, the noise reduction factor is $R = 0.955 \pm 0.002$, from which we can estimate the effective quantum efficiency in this asymmetric configuration to be $\eta = (7.0 \pm 0.2)\%$. This value, slightly lower than that in the symmetric configuration ($i.e.$ 8.5$\%$), accounts for additional losses introduced in both arms as a result of increasing the distance from the nonlinear crystal to allow for the insertion of optical elements in the signal arm.

Concerning the two noise signals, we consider two quasi-single-mode thermal states with mean values $\langle m_{\rm N}^{(0)} \rangle = 0.176 \pm 0.001$ encoding the logical bit value 0, and $\langle m_{\rm N}^{(1)} \rangle = 0.381 \pm 0.005$ encoding the logical bit value 1. Their photon-number distributions are shown together in Fig.~\ref{Rwithnoise}(b). 
When either noise signal is superimposed on the lossy transmission channel, the value of $R$ measured by Bob increases compared to the noiseless case [see Eq. (\ref{noisylossyR})], though it remains below 1, as shown in Fig.~\ref{Rwithnoise}(c). Specifically, the superposition of the two noise signals yields two values---$R^{(0)} = 0.962 \pm 0.002$ and $R^{(1)} = 0.975 \pm 0.002$, respectively---that are different within 2$\sigma$.

It is worth noting that the fact that all these values are below 1, but not by much, makes the discrimination strategy particularly intriguing: under these conditions, a potential intercept-resend attack by Eve can easily raise the value of $R$ above 1 or, at least, above a given threshold of uncertainty, thus providing a clear and readily applicable criterion for detecting the attack.
From the experimental point of view, these conditions can be easily attained by properly choosing the wavelength of the produced TWB or the model of SiPMs, \textit{i.e.}, by suitably tuning the detection efficieny $\eta$ [see Eq. (\ref{noisylossyR})].

\subsection{Performance of the discrimination strategy}
In the proposed communication protocol, we assume that Alice prepares the TWB, chooses the binary sequence of noise signals to superimpose on the signal arm, and sends it to Bob. While Alice measures the light in the idler arm, Bob does the same in the signal arm and compares his results with Alice's, evaluating either $\langle m_s \rangle$ or $R$ to decode the received bits. In this Section, we compare the performance of these two decoding strategies, one based on a classical feature, $\langle m_s \rangle$, and the other on a nonclassical one, $R$.

\begin{figure}[!t]
\centering
\includegraphics[width=\columnwidth]{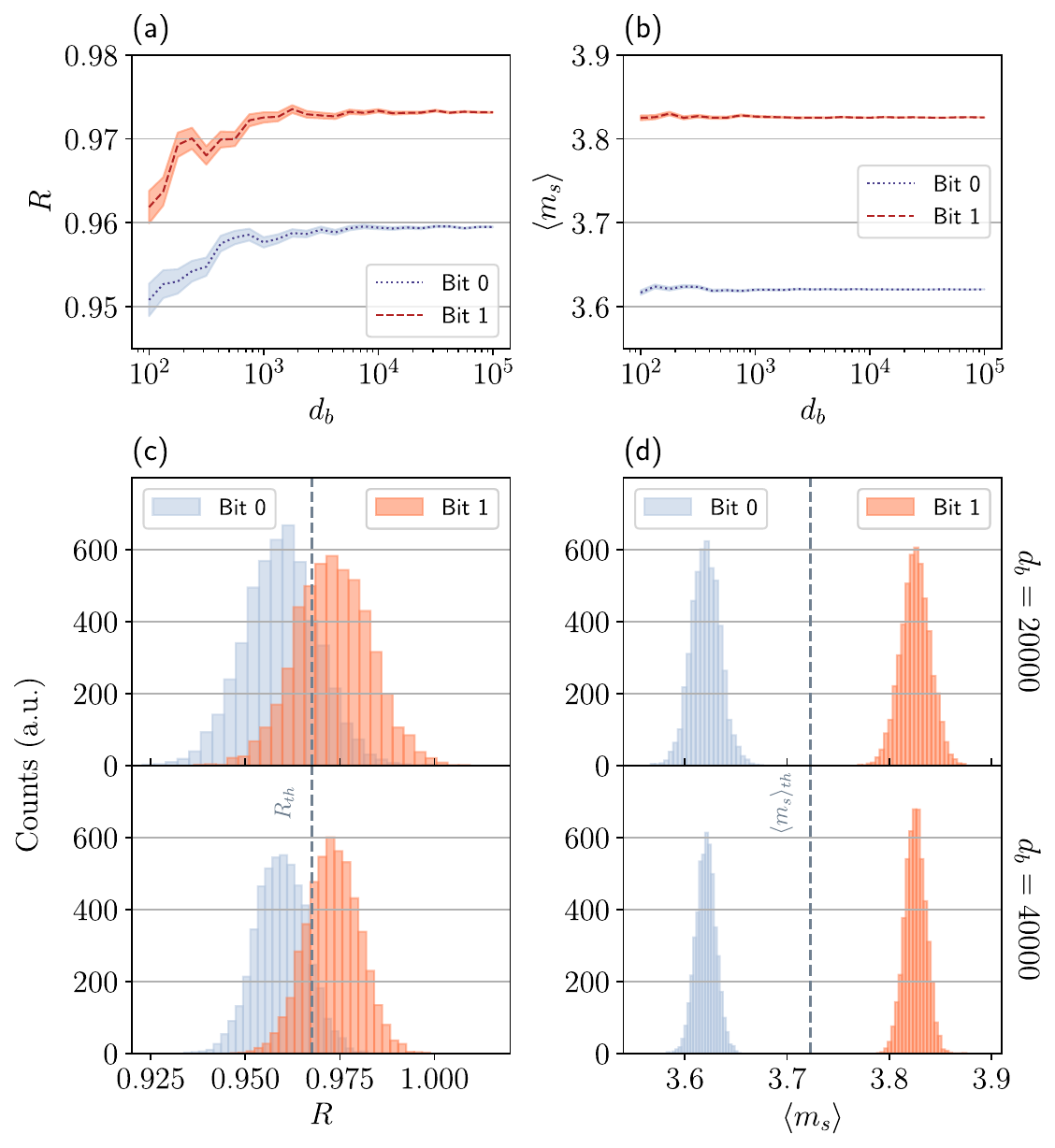}
\caption{Effects of the batch size on the estimation of $R$ and $\langle m_s \rangle$. Mean value and associated error of the noise reduction factor $R$ (a) and of the mean number of detected photons in the signal arm $\langle m_s \rangle$ (b) as a function of the batch size $d_b$. Histogram of the values of $R$ (c) and of $\langle m_s \rangle$ (d) for $d_{b} = 20000$ (top) and $d_b=40000$ (bottom). In (c,d) the vertical dashed line denotes the discrimination threshold $R_{th}=0.968$ and  $\langle m_s \rangle_{th}=3.72$, respectively. Results are obtained from $5000$ batches drawn from the original datasets by bootstrap procedure.}
\label{fig:qkd_prot_hist_R_m}
\end{figure}

For this analysis, a remark is in order. The $10^5$ shots we measured for each dataset (lossy TWB and noise signal with $\langle m_{\rm N}^{(0)} \rangle$ and $\langle m_{\rm N}^{(1)} \rangle$) are not sufficient to properly test the performance of the discrimination strategy, because a reliable estimate of $\langle m_s \rangle$ and $R$ requires at least $O(10^4)$ measurements, as shown in Fig. \ref{fig:qkd_prot_hist_R_m}(a,b). This is particularly necessary for evaluating the noise reduction factor, which depends on second-order moments of statistical distributions [see Eq. (\ref{Rdetphot})]. To overcome this limitation, we apply the bootstrap procedure \cite{efron,boot} to the original experimental datasets from which we draw samples. Each batch we draw represents a possible realization of the experiment. In the following, we evaluate and discuss the performance of the discrimination strategies based either on $R$ or on $\langle m_s \rangle$ using $5000$ batches for representative values of the batch size $d_b$.

To illustrate the different discrimination capabilities of $R$ and $\langle m_s \rangle$, it is instructive to focus on the distribution of their values 
for two representative batch sizes $d_b=20000,40000$, as shown in Fig. \ref{fig:qkd_prot_hist_R_m}(c,d).
Letting $x=R,\langle m_s \rangle$ be the discrimination parameter, here we define the discrimination threshold as $x_{\rm th} \equiv (x^{(0)}+x^{(1)})/2$, where $x^{(b=0,1)}$ is the value obtained from the original datasets (values are reported in Sec. \ref{sec:results_twb}).
Discriminating the superimposed noise signal ($i.e.$ decoding the bits) via $R$ leads to misclassification errors: the two histograms representing bits 0 and 1 overlap [Fig. \ref{fig:qkd_prot_hist_R_m}(c)]. In contrast, the outstanding PNR capabilities of SiPMs enable the exact discrimination when using $\langle m_s \rangle$: there is no overlap of the two histograms representing bits 0 and 1 [Fig. \ref{fig:qkd_prot_hist_R_m}(d)].

\begin{figure}[!t]
\centering
\includegraphics[width=\columnwidth]{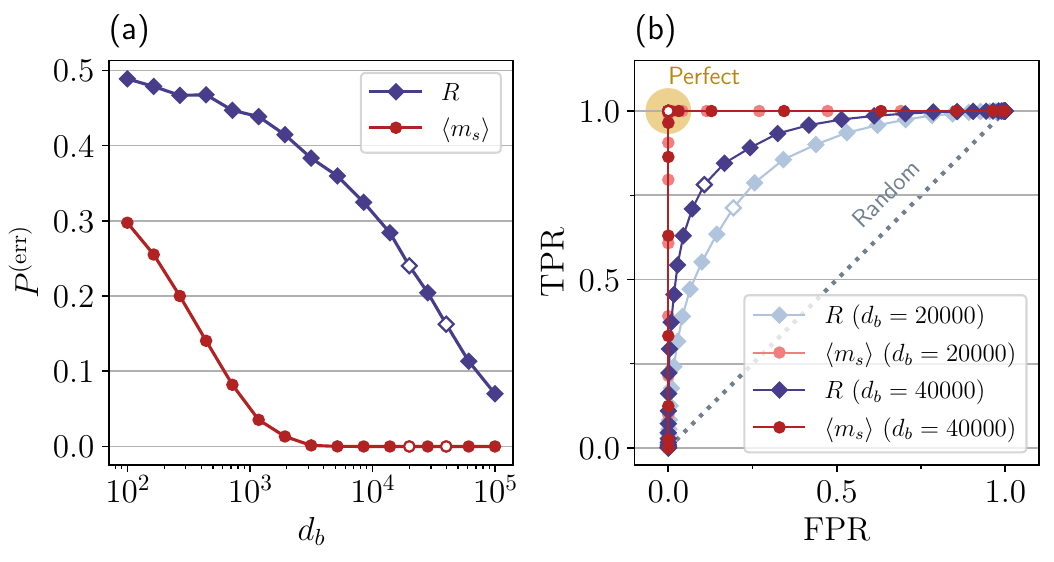}
\caption{Effects of the batch size and threshold on the discrimination of bits. (a) Error probability (misclassification) as a function of the batch size $d_b$ when using $R$ (blue diamonds) or $\langle m_s \rangle$ (red circles) to discriminate bits 0 or 1 for a given threshold (see Fig. \ref{fig:qkd_prot_hist_R_m}). (b) ROC of each discrimination strategy: True positive rate (TPR) as a function of the false positive rate (FPR) for two representative batch sizes $d_b=20000, 40000$. The perfect classifier corresponds to (FPR=0,TPR=1) (yellow shadow), while the random classifier to the line TPR=FPR (grey dotted line). The empty markers denote in (a) data associated with $d_b = 20000,40000$, and in (b) those associated with threshold values $R_{th}=0.968$ and  $\langle m_s \rangle_{th}=3.72$. Results in (a)-(b) are obtained from $5000$ batches drawn from the original dataset by bootstrap procedure.}
\label{fig:qkd_prot_class}
\end{figure}

As discussed in Section~\ref{sec_bin_class_pbm}, to better understand how the batch size $d_b$ affects the discrimination of bits, we consider the following figures of merit at varying $d_b$:
(i) The error probability $P_x^{\rm (err)}$ defined in Eq.~(\ref{errprob}), according to which a bit 0 is misclassified as bit 1 or vice versa, (ii) the ROC curve, and (iii) the associated AUC.
(i) The error probability in Fig.~\ref{fig:qkd_prot_class}(a) clearly shows how the discrimination via $\langle m_s \rangle$ outperforms that via $R$ even at low $d_b$, where discrimination via $R$ is basically random guessing. Focusing on $d_b \sim 10^4$, we have $P_{\hbox{\tiny $\langle m_s \rangle$}}^{\rm (err)} \approx 0$, while $P_{\hbox{\tiny $R$}}^{\rm (err)} \approx 16\%$ for $d_b = 40000$.
(ii) The ROC curves in Fig.~\ref{fig:qkd_prot_class}(b) reveal that for $d_b=20000, 40000$ both discrimination strategies perform better than random guessing, though with significant differences. The ROC curves related to $\langle m_s \rangle$ for the two $d_b$ overlap and perfect discrimination is enabled, $({\rm FPR},{\rm TPR})=(0,1)$, for various threshold values. Instead, the ROC curve related to $R$ improves upon increasing the batch size, but this discrimination strategy requires a fine-tuning of the threshold value, and yet it performs worse than the strategy based on the mean value.
(iii) Analogous conclusions can be drawn by evaluating the area under the ROC curve: ${\rm AUC}\approx 1$ (perfect classifier) when discriminating via $\langle m_s \rangle$ for both the batch sizes considered, while ${\rm AUC} \approx 0.843$ if $d_b=20000$ and ${\rm AUC}\approx 0.918$ if $d_b=40000$ when discriminating via $R$.

To summarize, reliably encoding a bit requires a 
batch size at least $d_b \sim O(10^4)$, in particular when using $R$ rather than $\langle m_s \rangle$ [see Fig. \ref{fig:qkd_prot_hist_R_m}(a,b)]. This is also reflected in the error probability for decoding a bit (\textit{i.e.}, discriminating the noise signals) [see Fig. \ref{fig:qkd_prot_class}(a)]. As evidenced by the analysis of the ROC curve [see Fig. \ref{fig:qkd_prot_class}(b)] and the AUC, the $\langle m_s \rangle$-based discrimination strategy exactly distinguishes the noise signals without requiring a fine-tuning of the threshold, in contrast to the $R$-based strategy, which is prone to errors.

\begin{figure*}[!t]
\centering
\includegraphics[width=1\textwidth]{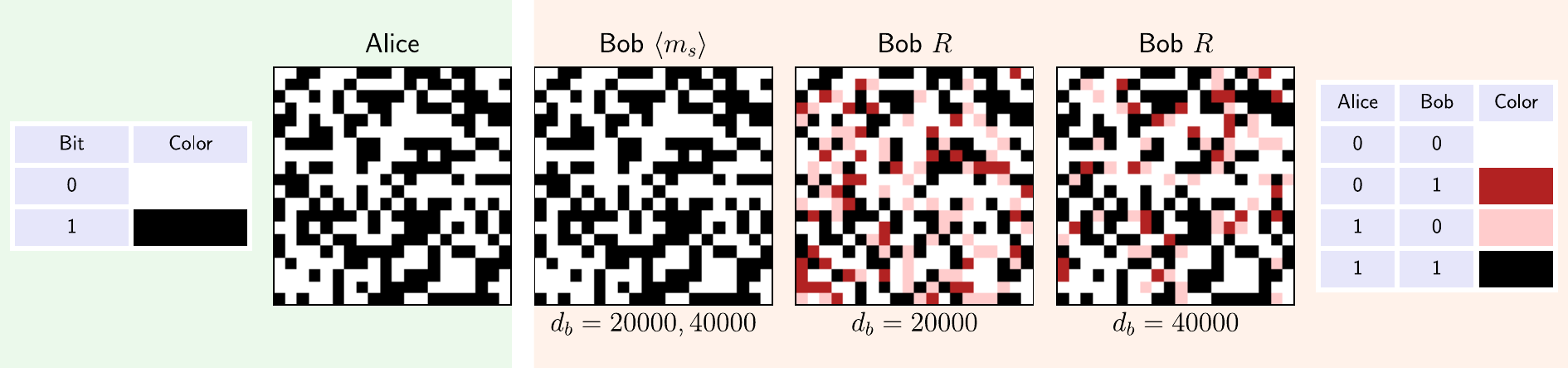}
\caption{Example of discrimination of a 400-bit key for two batch sizes $d_b=20000,40000$. The raw key is visualized as a 20×20 binary image, where each pixel corresponds to a single bit. Alice generates her key using a pseudo-random number generator, while Bob obtains his key by measuring either $\langle m_s \rangle$ or $R$ and discriminating bits according to the thresholds reported in Fig. \ref{fig:qkd_prot_class}. Errors in Bob's keys, highlighted in light- and dark-red, occur with rates determined from $5000$ batches drawn from the original dataset by bootstrap procedure.}
\label{fig:key_example}
\end{figure*}
\section{Proof-of-principle test}\label{sec_discussion}
In light of the results presented in Sec.~\ref{sec_results} for the two discrimination criteria, it is interesting to investigate the performance of each discrimination strategy. 
We point out that while $\langle m_s \rangle$-based discrimination strategy relies only on Bob's measurement outcomes, the $R$-based one requires Bob and Alice to share, over a different communication channel, their measurement outcomes with each other [see Eq. (\ref{Rdetphot})].
As a proof-of-principle test, we generate a binary string (Alice's raw key) with a pseudo-random number generator \cite{python}. The rates of true/false positive/negatives---estimated over $5000$ batches drawn from the original datasets---define the probabilities that govern the decoding of each bit in Bob's raw key. Fig. \ref{fig:key_example} shows an example of a 400-bit raw key decoded using the two discrimination strategies, for two batch sizes $d_b=20000,40000$.

As expected, Bob can exactly decode Alice's raw key using the $\langle m_s \rangle$-based discrimination strategy, while the raw key he obtains using the $R$-based strategy is affected by errors. In particular, the error probability decreases from $P_{\hbox{\tiny $R$}}^{\rm(err)}\approx 24\%$ to $\approx 16\%$ upon increasing the batch size from $d_b =20000$ to $40000$ [see also Fig. \ref{fig:qkd_prot_class}(a)]. In detail, the false positive/negative rates decreases from ${\rm FPR} \approx 19\%$ to $\approx 11\%$ and ${\rm FNR} \approx 29\%$ to $\approx 22\%$. Therefore, a bit-1 decoded as a bit-0 (\textit{i.e.}, a false negative) constitutes a misclassification error that occurs systematically more often than the opposite case, where a bit-0 is decoded as a bit-1 (\textit{i.e.}, a false positive). This asymmetry in error rates arises from the choice of the discrimination threshold $R_{th}$ in the present example. According to the analysis of the ROC curve, and recalling that ${\rm FNR} \equiv 1 - {\rm TPR}$,  the condition ${\rm FNR} =  {\rm FPR}$ corresponds to points lying along the anti-diagonal ${\rm TPR} = 1 - {\rm FPR}$. However, as illustrated in Fig. \ref{fig:qkd_prot_class}(b), the points associated with $R_{th}=0.968$ lie below the anti-diagonal, thus indicating ${\rm FNR} >  {\rm FPR}$ (see also Fig. \ref{fig:qkd_prot_hist_R_m}(c), which shows that the threshold $R_{th}$ lies closer to one distribution than to the other).
Since this protocol requires estimating $R$, each bit must be encoded using a batch size at least $d_b \sim O(10^4)$. In this regard, $d_b \approx 40000$ represents a suitable batch size to reliably estimate $R$ while keeping the number of shots per bit reasonably limited.

In light of these results, one might thus be tempted to choose the discrimination strategy based on $\langle m_s \rangle$ and to use a lower batch size, which in turn saves resources and can increase the bit rate, given the mild impact of batch size on discriminative power. However, the $\langle m_s \rangle$-based protocol is vulnerable to eavesdropping, \textit{e.g.}, to intercept-resend attacks, which can instead be readily detected by the values of $R$. 
Therefore, the hybrid discrimination strategy we propose is devised to leverage a classical property---the mean value of the detected light, $\langle m_s \rangle$---for bit discrimination, while harnessing a nonclassical feature---the noise reduction factor $R$ between the two arms of the TWB---to ensure security.
In practice, Bob constructs his raw key by decoding the received light using $\langle m_s \rangle$, and retains each bit only if the corresponding measurement satisfies $R < 1$ (nonclassical correlations) and the value of $R$ falls within a reasonable uncertainty range.

\begin{figure}[!t]
\centering
\includegraphics[width=\columnwidth]{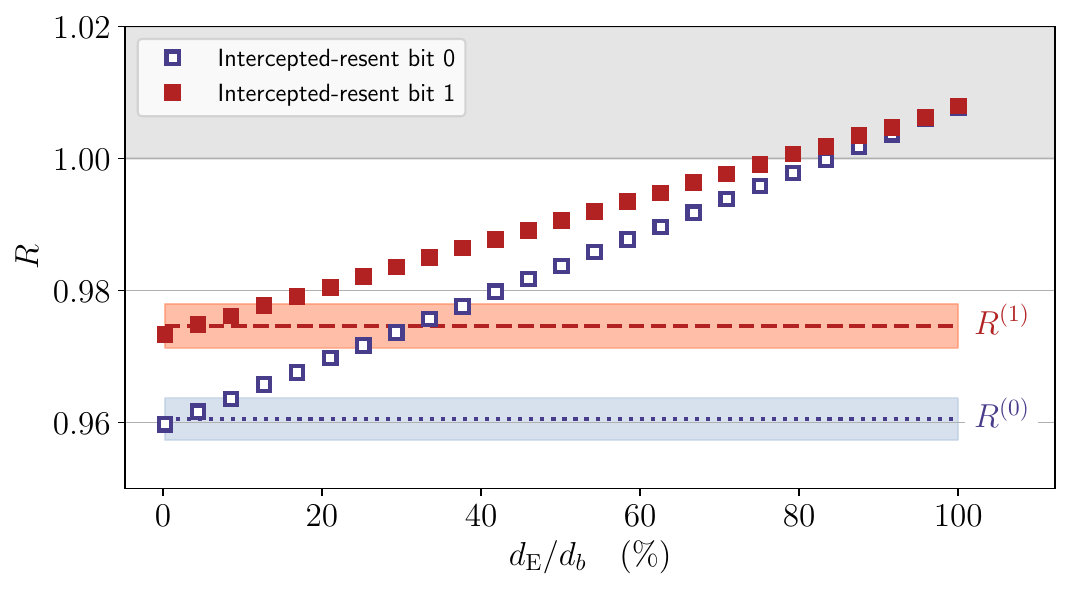}
\caption{An ideal intercept-resend attack can be revealed by the values of the noise reduction factor $R$. Eve intercepts $d_{\rm E}$ shots of a batch encoding a bit ($d_b=40000$) and resends, in place of the intercepted light, $d_{\rm E}$ shots belonging to a Poissonian light having the same mean value as that of the intercepted light. While the mean value measured by Bob is not affected by this attack (data not shown), the value of the noise reduction factor is. Blue color refers to bit 0, while red color refers to bit 1. The blue dotted line + blue highlighted band corresponds to $R^{(0)} \pm 2\sigma$, while the red dashed line + red highlighted band corresponds to $R^{(1)} \pm 2 \sigma$. The values of $R$ corresponding to the intercepted-resent bit 0 (1) are shown as blue open squares (full red squares). Results are averaged over $5000$ realizations of the intercept-resend attack.}
\label{fig:eavesdropping}
\end{figure}

To better investigate the feasibility of the strategy, we simulate an eavesdropping attack by assuming that Eve intercepts part of a batch encoding either 0 or 1, and re-sends a signal with the same mean value as the stolen portion of light. In doing so, Eve replaces a portion of the shots in the signal arm (Bob)---originally nonclassically correlated with those in the idler arm (Alice)---with others from a different light source, which is uncorrelated with Alice's. This degradation of nonclassical correlations due to the intercept-resend attack is therefore expected to increase the noise reduction factor [see Eq. (\ref{noisylossyR})]. In particular, we consider a synthetic signal generated according to a Poissonian distribution, since this is the statistical distribution that better approximates the distribution of the signal received by Bob in the absence of eavesdropping.
The results of this intercept-resend attack are shown in Fig.~\ref{fig:eavesdropping}, where the noise reduction factor is shown as a function of the ratio $d_{\rm E}/d_{b}$ between the batch size changed by Eve and that encoding either 0 or 1. The linear dependence of $R$ on $d_{\rm E}/d_{b}$ can be proved analogously to the derivation of Eq. (\ref{noisylossyR}). The highlighted bands indicate the uncertainty corresponding to $\pm 2 \sigma$.
It can be noticed that for bit 0 the value of $R$ exceeds $R^{(0)}+2\sigma$ and $1$ at $d_{\rm E}/d_{b}\approx 9\%$ and $\approx 84\%$, respectively. Instead, for bit 1 it exceeds $R^{(1)}+2\sigma$ and $1$ at $d_{\rm E}/d_{b}\approx 14\%$ and $\approx 78\%$, respectively. This means that Bob can easily detect the presence of Eve by simply checking whether the values of $R$ are larger than $2 \sigma$. On the contrary, values $R>1$ are more difficult to observe, at least when Eve, as in the present example, is equipped with an ideal detector that allows perfect estimation of the mean value of the intercepted signal. At the same time, this proof-of-principle test does not take into account possible optimization strategies, such as practical reconciliation or privacy amplification, devised by Bob to improve his receiver.

Moreover, in view of a realistic application of the communication protocol, the scheme can be implemented at telecom wavelengths \cite{kumar,eckstein}, while leaving the detection in the visible region and using a sum-frequency generation process \cite{bai,chae} before the detection stage. Work on this phase is ongoing precisely on a SiPM-based detection scheme.

\section{Conclusions}\label{sec_conclusions}
In this work, we investigated a hybrid state discrimination strategy in the mesoscopic intensity domain, where information is encoded in the photon number of two single-mode thermal states. The proposed method employs a receiver that uses PNR detectors, while the communication channel exploits mesoscopic TWB states.
The PNR capability provides direct access to the mean photon number, which is essential for effective discrimination. Simultaneously, the use of nonclassically correlated TWB states, characterized through the noise reduction factor, enables the detection of eavesdropping attacks, thereby enhancing communication security.
Our analysis focuses on the error probability and the ROC curve, both evaluated as functions of the batch size of transmitted signals.

The results demonstrate the existence of a batch-size threshold above which the chosen discrimination metric becomes reliable. Specifically, discrimination based on the mean number of photons achieves lower error rates and requires fewer measurements, but does not provide inherent security. Conversely, discrimination based on the noise reduction factor $R$ can reveal eavesdropping activity, although it is more error-prone and demands a larger number of measurements per bit. Furthermore, the analysis of the ROC curve offers insights into the nature of errors occurring in the communication channel.

The overall robustness of the proposed discrimination strategy paves the way to further enhancements, particularly in the encoding scheme. For example, one promising direction would be to adopt an alphabet with more than two symbols \cite{becerra}, encoded either in single-mode thermal states with distinct mean values or in superthermal states \cite{pla23}.

\section*{Data Availability Statement}
The data that support the findings of this study are available from the corresponding author upon reasonable request.

\section*{Acknowledgments}
We thank Silvia Cassina (University of Insubria) for fruitful discussions. A.~A. acknowledges support from Grant No. PNRR D.D.M.M. 737/2021. Scientific support from CRIETT centre of University of Insubria (instrument code: MAC27) is greatly acknowledged. L.~R. acknowledges support from INFN through the project ``QUANTUM''.

\section*{Conflict of interest}
The authors have no conflicts to disclose.

\section*{Author contributions}
\textbf{Luca Razzoli:} Data curation (equal); Formal analysis (lead); Methodology (equal); Writing – Original draft (equal); Review and editing (equal). \textbf{Alex Pozzoli:} Investigation (equal); Software (lead); Data curation (equal); Review and editing (equal). \textbf{Alessia Allevi:} Conceptualization (lead); Investigation (lead); Supervision (lead); Methodology (equal); Funding acquisition (lead); Writing – Original draft (equal); Review and editing (equal).

\end{document}